\documentclass[]{iopart}

\def\insertplot#1#2#3#4#5#6#7{
\vskip 10pt\nobreak\hbox to \hsize{\hss\dimen0=#3in\hbox to #6\dimen0{%
\dimen0=#2in\vbox to #6\dimen0{\vss
	    \special{ps: plotfile #1}
	    \special{ps::[end]
	      PGPLOT restore
	      }
	      }\hss}\hss}\vskip 10pt}

\begin{document}
{Similar to version that will be published by IOP in 2008, Physica Scripta T, 
\newline
(Special Issue on Nobel Symposium 135: Physics of Planetary Systems)}

\title[Disk Dispersal and Planet Formation Time Scales]{Disk Dispersal and Planet Formation Time Scales}

\author{Lynne A. Hillenbrand}

\address{California Institute of Technology;
Mail Code 105-24; Pasadena CA 91125; USA}
\ead{lah@astro.caltech.edu}
\begin{abstract}
Well before the existence of exo-solar systems was confirmed,
it was accepted knowledge that most -- if not all -- stars possess 
circumstellar material during the first one-to-several million years 
of their pre-main sequence lives, and thus that they
commonly have the potential to form planets.  Here I summarize 
current understanding regarding the evolution of proto-planetary 
dust and gas disks,
emphasizing the diversity in evolutionary paths. 
\end{abstract}

\section{Introduction}
From analysis of the chemical record as traced through meteoritic material,
we can infer the detailed history of the formation and early evolution
of our planetary system (e.g. Ehrenfreund, this volume).
Since discovery of the first exo-solar system planet more than a decade ago,
it has been understood that planetary systems around other stars could have 
properties very different from those characterizing the familiar Solar System. 
Further, the biases and systematics that are inherent in the many exo-solar system planet detection  techniques, plus the current technological limitations,
 mean that the discovery of planetary systems 
resembling our own -- in detail -- remains for the future.  

However, the great diversity in the measured orbital and internal properties 
among known ``exo-planets" suggests that it 
would not be surprising were there similar diversity in the properties of 
the circumstellar disks out of which such planets form. 
Indeed there is a wide range in inferred
disk size, disk mass, disk geometry/structure, and disk composition/chemistry, 
as touched upon by other authors in this volume 
(e.g. Henning, Natta, Dullemond, Aikawa).  While the initial conditions 
for planet formation do appear diverse, typical early-stage properties are:
M$_{disk}$ = 0.005 M$_\odot$ 
(e.g. Osterloh \& Beckwith 1995; Andrews \& Williams 2005, 2007); 
M$_{disk}$/M$_{star}$ =  1-10\%;
dM$_{acc}$/dt = 3$\times$ 10$^{-9}$ M$_\odot$/yr 
(e.g. Gullbring et al. 1998; White \& Basri 2003;
Muzerolle et al. 2003); and R$_{disk}$ = 10-100 AU (e.g. Dutrey et al. 1996).  
We note that the dispersion can be orders of magnitude 
for some disk properties.  

How does gas and dust that are initially well mixed and smoothly varying
with radius and height turn into dynamically and compositionally diverse planetary
systems?  Here, we discuss the process of disk dissipation and {\it presumed} 
planet building, with particular attention to the current 
observational constraints on relevant time scales.  
We focus on stellar populations as a whole, rather than describing 
the many individual cases of disk evolution ``in action," 
e.g. lines of evidence for grain growth
which are covered by other authors in this volume.
Theory has advanced in recent years
to the point of making specific predictions for the evolution of
quantities that, in principle, can be observationally constrained: 
surface density with radius (the dissipation is inside/out in some models);
total disk mass (predicted as nearly constant in some models);
rate of disk accretion (ceases entirely in some models);
gas-to-dust ratio (decreases from $\sim$100 to $<$1\%); 
grain size distribution (expect increasing mean size); 
and chemical composition (set by response to x-ray, EUV, FUV, 
and optical photon heating from the star).  While some of 
these measures are still beyond the realm of observational constraint, 
other tests having statistical significance are possible now.  
A potential ``second parameter" effect is
the local environment of the young star/disk system, and whether it is
dynamically or radiatively important to the evolution of the disk 
(see Adams, this volume for discussion of cluster effects and
Monin et al., 2007 for a review of stellar and substellar
 companion properties and their effects). 

\section{Planet Building}

Observational probes of the planet formation process are directed towards
understanding the evolution of circumstellar dust and gas properties.
The basic processes on the dust side are those 
of decoupling from the gas, drift and mid-plane settling, 
coagulation and growth into larger grains, and consequent growth to macroscopic 
bodies termed ``pebbles" then ``planetesimals," which can continue to 
become ``oligarchs", and eventually terrestrial sized planets.  
These processes may be traced by studying in large samples of disks e.g. 
spectral diagnostics of grain size, dust opacity vs wavelength, 
dust mass at optically thin wavelengths, disk structure as indicated by direct
imaging or inflections in spectral energy distributions, etc. 
On the gas side, there is a maze of chemical evolution before and then 
during accretion of the gas -- if giant planets are indeed formed -- 
onto massive cores that were themselves accreted via the dust evolution 
processes just described.  Tracers of gas disk evolution include 
probes of hot atomic gas accreting onto the star, 
numerous atomic and molecular gaseous species in more quiescent 
yet still warm regions of the disk, and measures of total gas mass via 
optically thin line emission.  

The planet building processes all occur against the backdrop of viscous disk 
evolution (Hartmann et al. 1998; Takeuchi et al. 2005; Alexander \& Armitage 2007), 
outflow via stellar/disk winds (Pudritz et al. 2007),
ionization/photoevaporation (Hollenbach et al. 1994; Gorti \& Hollenbach 2008),
and radiative blowout of small grains once the disk becomes optically thin. 
Various time scales are involved, ranging from those relevant
to radiative transfer and chemistry to those which scale with the dynamical 
time in the disk.
Here, we discuss mostly the dust disk evolution, since observational
constraints on gas disk evolution are more limited at present. Both remain
hindered to some extent by observational sensitivity, despite significant
advancement in recent years.  

How long will known disks last?
One relevant experiment is to consider the well-studied young stellar 
population of the Taurus-Auriga region for which all of the following are
available: 1) detailed spectral
energy distributions, including data from the Spitzer Space
Telescope; 2) disk masses from millimeter measurements,
and 3) mass accretion rates from inner disk to star, 
measured via either high dispersion spectroscopic measurements of ``veiling" 
or lower dispersion direct detection of the Balmer continuum. 
A simple division of M$_{star}$/(dM$_{acc}/dt)$ yields a time scale -- ranging from a few
Myr to roughly a Gyr -- for the accretion of material on to the central star.
That these times are much longer, in the mean, than the inferred stellar ages 
indicates that the accretion rates must have been much higher 
in the past in order to build up the stellar mass to its present value.  
This point has been made before in the literature, repeatedly.  

The complementary division of M$_{dust}$/(dM$_{acc}/dt)$ 
also yields a time scale, that for the disk to dissipate under the
assumption that all of the material currently residing in the disk eventually
winds up on the star.  The inferred times are factors of several longer than
the stellar ages estimated at 1-2 Myr, suggesting that the disks 
will last well into the future.  Given the
simplistic assumptions regarding dust opacities used in estimating 
dust masses, they are likely underestimated, perhaps by an order
of magnitude (e.g. Draine, 2006), and this would strengthen our argument
regarding the potential for ``long-lived" dust disks.
When considered relative to typical theoretical time scales for
planetary core formation and gas accretion, {\it all young disks with
substantial dust and gas thus appear to have the potential to form planets} 
and can be considered proto-planetary.

How long does disk dissipation take, whether due to planetary formation
or other processes?  It has been known for some time
(e.g. Skrutskie et al. 1990 working with IRAS, Nordh et al. 1996
with ISO) and confirmed with more recent data
 (e.g. Hartmann et al. 2005 and Furlan et al. 2006 with Spitzer) 
that 3-25 $\mu$m mid-infrared colors and SED slopes of young low mass stars 
(also known as T Tauri stars) in a single cluster segregate into two groups.  
They are interpreted as the stars with disks and the stars without disks. 
As the wavelength considered is decreased, blurring between the 
excess/disked and the non-excess/non-disked samples increases.  
Of note is that
similar bifurcation in mid-infrared colors is seen not only among 
very young 1-2 Myr populations but also in somewhat older 5-10 Myr old 
populations such as the $\eta$ Cha (Megeath et al. 2005) and 
TW Hya (Low et al. 2005) small associations. It is always present.

That there are no, or very few, objects of intermediate color or spectral slope
found at mid-infrared and longer wavelengths among stellar populations of 
nominally the same age has been used to argue 
(e.g. Simon \& Prato 1995; Wolk \& Walter 1996)
that the transition time from optically thick to optically thin disks is
only a few hundred thousand years, i.e. that the disk dissipation process
is quite rapid -- once it starts.  Bertout et al. 2007 and
Hartigan et al. 1995, however, present evidence to the contrary that so-called 
``CTTS" and ``WTTS", i.e. Classical accreting and Weak-/non-accreting
T Tauri Stars,  
are distinguishable in their luminosity ({\it sic} age) distributions.

The key to understanding the apparent diversity in the disk dissipation
time scale may be found by considering whether the differences are due to
those in the initial conditions from which the evolutionary process occurs, 
or to differences in the process itself or its duration.
Notably, self-similar viscous disk evolution
models (e.g. Hartmann et al. 1998) do not yield the rapid transition between
strong and weak/no disks that is observed.  Models invoking inside-out
ionization/photoevaporation (e.g.  Alexander \& Armitage 2007) or 
rapid grain growth and planetesimal formation (e.g. Brauer,
Dullemond, \& Henning 2008) may.  

\section{Transitional Disks}

A category of disks which has been identified for close to twenty years but
has only recently become well-defined, is the so-called ``transition" disk
sample.   These objects make up a very small fraction of the total disk population.  The term ``transitional" is commonly used to describe several
different categories of disks that are perhaps indicative of an
``evolved" nature.  There are those disks
having less strong spectral energy distributions overall, 
relative to their counterparts.  There are those disks having low 
dust content in the inner ($<$1-10 AU) regions based on 
no or weak near- and mid-infrared excesses,  yet strong 
excess emission and quite diverse spectral energy distributions, 
at longer mid-infrared 
(e.g. Furlan et al. 2006; Watson et al. 2007) through sub-mm/mm wavelengths. 

The latter objects have a range of inferred dust masses that is typical
of the disked T Tauri population.  However,  
inner disk gas content, where measurements are available, 
is low with gas surface densities $<$1-2 g/cm$^2$ and 
gas:dust ratios 250-1000 inferred at 0.5-1 AU 
(Saylk, Blake \& Brown, 2007; Herczeg et al. 2007).
This is consistent with the much lower-than-average accretion rates 
inferred from optical and ultra-violet diagnostics. 
According to the M$_{dust}$/(dM$_{acc}/dt)$ metric 
applied above, the ``transition" disk sources are projected to be
exceptionally long lived accretion systems if their future evolution is determined
entirely by viscous dissipation.  However, dust/gas removal may occur
by other means, for example enhanced photoevaporation given that
the inner disk is already cleared and the outer disk thus
more directly illuminated by stellar photons. 

The list of candidate ``transition" disks is growing rapidly based on 
sensitive Spitzer data.  In well-studied star forming regions, Spitzer has
confirmed that many objects with only upper limits to their mid-infrared 
fluxes measured by IRAS, indeed have photospheric spectral energy 
distributions with an upturn at wavelengths longer than 
5-20 $\mu$m. In other less well-studied regions, 
Spitzer by itself has characterized the entire strong/transitional/weak
or non-disk population (e.g. Sicilia-Aguilar et al. 2006ab).

The relative paucity of ``transitional" objects 
has long been used to argue that the phase from optically thick 
accretion to optically thin dust- and gas-poor disks is short-lived.  
One interesting question to ask is what are the ages of {\it the known}
transitional systems, 
and how do they compare to those of the typical cluster or group 
member?  We find that there is {\it no difference} in the mean, median, 
or distribution (via K-S test) of ages for the two
populations in Taurus-Auriga.  This finding is related to
the apparent conundrum of CTTS and WTTS mixing 
in the Hertzsprung-Russell diagram (HRD), noted earlier.

It is thus the case that stars of apparently - if not nominally - 
the same age can have very different disks.  
This diversity is noted from consideration of bulk disk properties 
(presence/absence, total mass, size, accretion rate, etc.) 
as well as in the many details of 
dust grain size and composition that are available from surface layer
spectroscopy or thermal emission from the near the mid-plane 
(see e.g. Natta, this volume).

\section{Stellar Ages}

This brings us to discussion of stellar ages.
While there are a number of different empirically calibrated
stellar age estimation techniques for stars between $\sim$100 Myr 
and $\sim$10 Gyr old (e.g. Mamajek \& Hillenbrand 2008), 
for young stars, we are stuck essentially with the HRD
as the only tool to determine stellar ages.  I offer some cautionary words about 
what we can and what we should not believe about stellar ages 
inferred from individual measurements of log $L/L_\odot$ and log $T_{eff}$
for young stars.  

First, there are systematic concerns.  For example, several 5-10\% precise values
of the distance to the Orion Nebula Cluster have become available very recently
(Hirota et al. 2007, Jeffries 2007; 
Sandstrom et al. 2007; Menten et al. 2008), indicating that all previous
interpretations of the HRD for this cluster have suffered from systematic
overestimate of the stellar luminosities and hence underestimate
of the stellar ages.  Another systematic effect is that of unresolved binarity,
the implications of which on stellar luminosity estimates
given the apparent trends with primary mass in component mass ratios 
and separation distributions, remains poorly characterized empirically
-- even for very well studied clusters.

A further concern of systematic nature is that pre-main sequence evolutionary 
tracks vary substantially between theory groups.  
Comparison among available tracks reveals a trend in inferred stellar age 
that is relatively flat (reflecting consistency between the various sets
of evolutionary tracks) for earlier type F and G stars, 
but increasing to about 0.75 dex age differences 
(indicating strong variation among theory groups) 
for later type K and M young stars, as  
demonstrated in Hillenbrand, Bauermeister \& White 2008.  
In addition to model-to-model systematics, 
it seems that {\it all} currently available sets of tracks 
under-predict stellar masses by 30-50\% (Hillenbrand \& White 2004)
while simultaneously under-predicting low mass stellar ages by 30-100\% 
and over-predicting high mass stellar ages by 20-100\%, 
(Hillenbrand, Bauermeister \& White 2009).  Finally, from comparison of
presumably co-eval populations in pre-main sequence clusters, it is concluded that
the higher mass stars systematically appear older than the lower mass stars 
in the same cluster, regardless of adopted tracks or mean cluster age.  
Although observers are generally grateful to have the opportunity 
to impose theoretical interpretation on their data,
the ensemble findings suggest that detailed work
in stellar astrophysical theory is still needed, as is guidance from
dynamical mass measurements across the pre-main sequence.

Second, there are random errors having to do with the accuracy 
of observationally determined quantities.  These errors act to broaden
luminosity dispersion, which is often (quite erroneously) interpreted as 
evidence for true age dispersion.  Sources of random error include 
both astrophysical noise such as photometric variability, 
stellar/disk ``activity," and observational noise such as pure Poisson error
in the measurements along with conditions such as source crowding or 
high background and other non-photospheric emission specific to certain
young regions.

How much confidence can we place in stellar ages and hence inferred 
evolution of other physical variables based upon them? 
At present, a conservative estimate
is that ages are accurate to factors of no better than $\sim$2-3, 
including both systematic and random uncertainties.
Prudence thus dictates caution regarding 
the inference of e.g. star formation histories in molecular clouds and
cloud-free stellar associations, as well as  
in assessment of evolutionary time scales for
e.g. circumstellar disks or stellar angular momentum. 

Beyond absolute calibration of mean ages, we must ask whether all stars 
in a particular stellar cluster or association 
have the same age, or if there is evidence for age dispersion 
among cluster members (e.g. Tout, Livo, Bonnell, 1999). 
We (e.g. Hillenbrand, Bauermeister, White 2008) have been running
monte carlo simulations to test whether observed luminosity distributions 
are consistent with error distributions or, perhaps,
indicative of true age spreads. 
The evidence at this point suggests that the vast majority of cluster 
stars are consistent within the errors with having the same age
i.e. there are no discernible age spreads representing the bulk 
of young stellar populations.  Admittedly, there {\it are} some stars out 
on the tails of the luminosity distributions 
that seem hard to explain unless they
are individually suffering some large unidentified source of error
(e.g. Slesnick, Hillenbrand, \& Carpenter, 2008). 
St34 in Taurus, which appears 
on the old side, or PDS66 in Upper Sco which appears on the young side,
are well-known examples of such objects which remain enigmatic at present.

\section{Current Understanding of Disk Dissipation}

Returning now to the question of the time scale for disk dispersal and
planet formation, we proceed by assembling: 
1) a set of stars which are known members of young associations or clusters,
2) optical and/or near-infrared information which allows us to locate them
in the theoretical HR diagram, and
3) a quantitative disk diagnostic.  
Associations and clusters are useful bins because they provide statistical
robustness, they offer samples having a range of stellar and substellar masses,
and they provide stars having the same formation environment, 
metallicity, and stellar age (probably).
The optical and near-infrared 
photometric/spectroscopic data assembly is fairly ``bread-and-butter" at
this point in astronomical history, 
requiring only time/effort and a careful assessment of errors.
For the last need of a disk diagnostic, we could
choose an optical depth indicator such as monochromatic
infrared excess at one or more wavelengths
wavelengths, a proxy of total disk mass, or an inferred 
disk accretion rate.  For reasons of relative abundance and uniformity, 
we opt here to make use of infrared excess as our diagnostic. 

Dust grains radiate over a broad range of wavelengths depending 
on their temperature and size.  We can measure excess emission
above a stellar photosphere due to absorbing and thermally re-emitting dust in the circumstellar environment.  Such excess (infrared) emission 
can be measured via empirical colors, or more appropriately and 
more accurately, from colors corrected for 
a theoretically or empirically determined stellar photospheric contribution.  
To gain a complete picture of the disk dissipation process, 
we want to measure the excess amplitudes over a wide range of wavelengths, 
ideally, and of course over a range of ages.  

Independent of the wavelength studied or the mean age of the sample, there is
ample evidence for a range of infrared excess properties observed among
young pre-main sequence
stars.  For example, color excess distributions (e.g. 8-24 $\mu$m and
3.6-8 $\mu$m from Spitzer, or K-L and H-K from the ground) 
span a wide range of values at the youngest ages.  Towards older ages,
however, the distributions narrow, and eventually reflect just the error 
distribution.  Hernandez et al. 2007, for example, show the frequency distribution 
of near- to mid-infrared spectral energy distribution slopes in several clusters 
$<$5 Myr of age. The distributions are double-peaked, indicative of the 
well-known bimodal behavior between ``disked" and ``non-disked" stars referred 
to above.  Figure 1, employing a different analysis technique, shows
reddening-corrected near-infrared color excesses over photospheric values, 
which are broadly distributed rather than double-peaked, and likely indicative 
of the range in accretion properties exhibited by young stars.  From overall
consideration of star forming regions having nominally different ages, 
a trend of decreasing dust disk (mean) frequency towards older (mean) ages 
is observed.  But the devil is in the details.

\begin{figure}
\insertplot{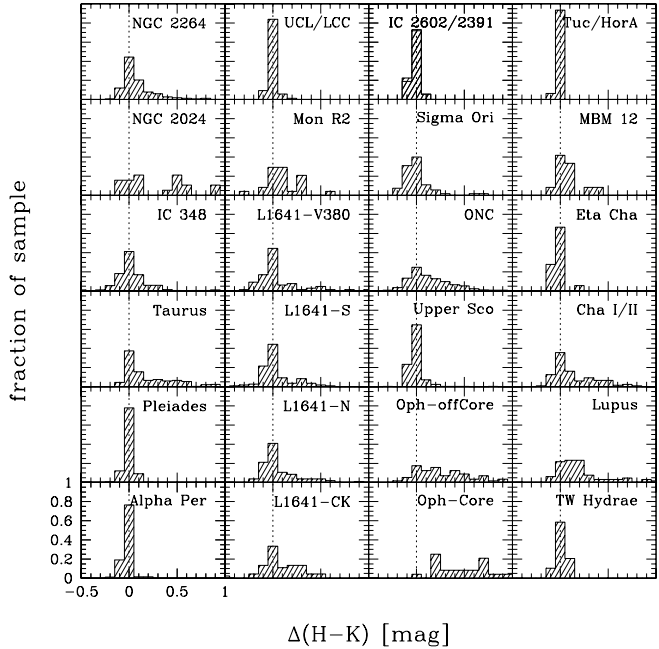}{6.5}{4.0}{1.0}{2.55}{0.45}{0}
\caption{Normalized frequency distribution of near infrared excesses
indicative (when positive)
of hot circumstellar dust associated with accretion processes
in $<$1-100 Myr old stars superposed (when negative) on the underlying
error distribution.  The younger clusters show a range of
near-infrared excess amplitudes (including zero excess) while the
older clusters are more strongly peaked at near-zero excess.}
\end{figure}

\subsection{Disk Dissipation Trends with Stellar Radius}

Short infrared wavelengths trace hot dust and thus
near-infrared excesses (JHK bands) 
sample material at $\sim$0.03-0.1 AU.  Longer wavelengths trace cooler dust 
with mid-infrared excesses (LMNQ bands or Spitzer IRAC camera) probing   
$\sim$0.2-10 AU and far-infrared and sub-mm/mm wavelengths typically 
$\sim$10-100 AU.  Stellar temperature/luminosity are also influential in
setting these ranges; the above numbers are relevant for a young ``solar-type" star.  
It should not be forgotten that there are significant radiative transfer effects 
introduced on top of this simple picture due to radial and vertical disk structure
as well as grain properties.  Regardless, in the innermost disk regions, 
a lack of 1-2 $\mu$m
excess observable from the ground implies $<10^{-5} M_\odot$ ($<3.5 M_{Earth}$)
in dust at the relevant temperature of $\sim$1000K. 
In the mid-disk range, lack of 8 and 24 $\mu$m excess with Spitzer 
implies $<1 M_{Earth}$ in $\sim$100 K dust. In the outer disk,
where the prevalent upper limits to sub-mm and mm excesses are less restrictive, 
only $<10^{-4} M_\odot$ ($<35 M_{Earth}$)
in 30-50K dust is implied by a non-detection.  

In the near infrared, measuring typically sub-micron grains,
we believe that the excesses are tied directly to the
accretion of material from the inner disk to the star.  Indeed, the
general decay of near-infrared excess with advancing stellar age over
1-10 Myr is well-reproduced by a similar decay in the frequency of H$\alpha$
emission for the same stars (Dahm, 2005; see also Mohanty et al. 2005,
Jayawardhana et al. 2006).  These arguments are buttressed by older evidence
in the literature for a 1:1 correlation among members 
of Taurus-Auriga between thermal emission from disks and gas emission
from accretion (Edwards et al. 1994).  Characteristic time scales are
a few Myr (see Hillenbrand, 2005 for more in depth discussion) and 
functional fits to the accretion diagnostics
of the form $time^{-1}$, $time^{-2}$, or $e^{-time}$ may be appropriate.

In the mid-infrared, measuring typically micron-sized grains, evidence 
is rapidly building from Spitzer (Silverstone et al. 2006, Lada et al. 2006,
Sicilia-Aguilar et al. 2006, Carpenter et al. 2006, Megeath et al. 2005,
Dahm \& Hillenbrand 2007)
but was also apparent from the ground (Mamajek et al. 2004) 
that a similar though perhaps slightly longer decay time can be inferred. 
In the sub-millimeter and millimeter, measuring potentially large grains 
up to tens or hundreds of microns in size, although no trends are available yet, 
there is a clear cliff of detectability that appears intimately tied 
to disk presence at shorter wavelengths. The data are consistent
with {\it all} CTTS having disks
with mass $>10^{-4} M_\odot$ and $<$10\% of WTTS
having substantial sub-mm disks (Andrews \& Williams 2005).   

There are, of course, some caveats to the above.  
One limitation is observational sensitivity, meaning that the weakest
or lowest mass disks are not detectable.  This is generally due either
to raw observational sensitivity limits, or to lack of 
calibration precision which effectively destroys capability to detect  
excesses within that uncertainty level.  A second limitation is in the
physical interpretation of employed disk diagnostics, 
i.e. how well we understand the physics leading to the excess flux
and how we correlate wavelength of the excess with temperature/location
of the emitting disk material. A final uncertainty, as discussed in detail above,
relates to the accuracies of stellar ages.

How do we interpret the current trends in disk dissipation vs wavelength
({\it sic}, disk radius)? We still -- and will for some time -- lack an unambiguous mapping
between observed spectral energy distributions and physical disks.  However,  
coming back to the so-called ``transitional disks," it is notable that 
the observations indicate morphologies suggestive of the dust becoming
optically thin first at shorter wavelengths and only later at longer 
wavelengths. This apparent ordering could be due to processes associated 
with physical disk draining (e.g. via accretion onto the central star, through
launching into a wind, or by photoevaporative processes that perhaps
progress from the inner disk to the outer disk) {\it or} to transformation
of disk material (e.g. the growth of small interstellar-like particles 
into grains larger than $\lambda/2\pi$ which are then not generally detectable 
at wavelength $\lambda$).

Indeed, an open question is whether
disk material dissipates at all radii simultaneously, or whether inner
disks disappear first, as holes develop on a viscous evolution,
dynamical, or photoevaporative time scale, and propagate outwards.  
The expected times are in all cases comparable to the dynamical
time scale - days in the inner disk and $\sim 10^5$ years in the outer disk. 
Although fast, these are perhaps not fast enough to produce 
the ``CTTS/WTTS switch" that is observed.  
One way of describing empirically the evolutionary process
may be via measurement of the slope of the infrared excess as it departs 
the stellar photosphere,
vs the wavelength at which the departure occurs (e.g.  Cieza et al. 2007).
It is proposed that grain growth implies clearing over a large range of
wavelengths near-simultaneously, while dynamical or photoevaporative effects
would proceed from the interior of the disk outward.  It is suggested that
these scenarios would have different tracks in such a diagram.

\subsection{Disk Dissipation Trends with Stellar Mass}

Another very clear trend from Spitzer data 
is the mass dependence of circumstellar disk dissipation.
Carpenter et al. 2006 and Dahm \& Hillenbrand 2007 show evidence 
in two different clusters of nominal age 5 Myr that the disks remain around only
the lowest mass (K and M) stars.  This is the first definitive evidence of the
long espoused notion that higher mass stars may lose their disks 
more rapidly. The finding is as one might infer from some combination 
of the larger radiation fields and the higher inferred mass accretion rates 
(Calvet et al. 2005, Garcia Lopez et al. 2006) for more massive stars, 
and is consistent with an interpretation that the mass dependent dissipation
trend is driven by initial conditions (Alexander \& Armitage 2005).
Further, the disks remaining in these two clusters (Upper Sco and NGC 2362), 
as well as others (such as $\lambda$ Ori, $\sigma$ Ori,
and Orion OB1) in even more recent literature, 
are weakened in strength or amplitude of the measured excess, relative
to the larger excess values observed towards stars 
in younger regions of similar mass.

How do we interpret the current trends in disk dissipation vs stellar age, 
as a function of stellar mass? Clearly, the fact that we can even
use terminology such as ``disk fraction" suggests that at any given
age, some stars have disks while others do not.  In other words, there is
dispersion - indeed, diversity -- in the time scale for disk
dispersal. This is true even among objects with apparently
identical properties otherwise (stellar mass, metallicity, star formation
environment).  Further, the mass dependence of both accretion properties 
and thermal dust emission processes is clear,  and also accompanied by
dispersion.  An open question is whether the observed trends 
should be interpreted as dispersion in initial disk properties, 
dispersion in the onset of some common disk evolutionary switch, 
or dispersion among individual objects in the relative importance of 
the various possible disk dissipation mechanisms.  

\section{Quantifying Diversity}

The wide dispersion in observed spectral energy 
distributions is especially prevalent at ages of $<$1-3 Myr and
perhaps indicative of dust disk geometry to first order
and of radiative transfer effects in detail.  
By $\sim$5 Myr the diversity settles down, with observed disks 
predominantly ``weak" and by $\sim$10 Myr most disks are
undetectable or nonexistent, with only a very few stars having ``strong" disks. 
What we really want to know, however, is not the evolution of observational
parameters -- e.g.  disk fraction, infrared excess amplitude, 
SED slope, $\lambda_{onset}$, optical veiling, emission line fluxes 
etc -- such as we have discussed or alluded to thus far,
or even the evolution of corresponding physical parameters
-- disk radial and vertical structure, 
total dust mass, grain size distribution, (d$M_{accretion}$/dt), 
gas mass etc.. Rather, we aim to understand a higher level question:
the frequency distribution of the lifetime of dust (as well as gaseous)
material above a certain mass, as a function of disk radius.  
How many such disks last only 0.1 Myr years, how many 1 Myr,
how many 5 Myr, 10 Myr, 20 Myr? 

This is approached, in principle,
via building the distribution of disk lifetimes as a function of wavelength.  
Ideally we want to go even further and understand the mean and dispersion 
in the evolution of physical quantities e.g. the disk surface density 
distribution, $\Sigma(r)$. Are such distributions gaussian
or do they exhibit long tails?  Does planet formation occur throughout
the distribution, or only within the long tails?  Is the circumstellar
evolutionary process different for stars of different mass, or for binary
vs single stars?

\section{Gas!}

We are unlikely to be able to discern in great detail what has happened to 
the ubiquitous early-stage circumstellar dust, via
studies of the dust itself. Instead, we might turn to studies of the gas, which
dominates the mass and therefore the disk dynamics.  In the
grain growth scenario for planet formation, the gas is likely to remain beyond
the dust disk lifetime and be available for continued disk accretion/outflow.
Conversely, in a disk clearing scenario accompanying planet formation, 
the gas likely disappears via the same or a similar mechanism to that
causing the dust removal. 
As in so many other areas of disk evolution, the less observationally
constrained gas is key to our astrophysical understanding.
Some of these same points have been made recently by Najita et al. 2007.

What of current gas constraints?  Emission in H$_2$ and CO 
has been detected from the ground 
(Weintraub et al. 2000; Bary et al. 2002, 2003;
Bitner et al. 2007; Salyk et al. 2007; Ramsay Howat \& Greaves, 2007) and 
measures warm-to-hot gas in disk surface layers or inner disk regions. 
With Spitzer, [Ne II] at 12.8 $\mu$m has been observed
(Pascucci et al. 2007; Lahuis et al. 2007)
and may actually measure the photo-evaporative flow
(Herczeg et al. 2007).  
Spitzer has also detected OH, H$_2$O and simple organic molecules
(Salyk et al. 2008, Carr \& Najita 2008).
Non-detections currently imply less than a few percent of $M_{Jupiter}$
remaining at the current age of the star, though samples are limited. 
Further gas studies, e.g. with the forthcoming
Herschel, are needed.  Of note is that the ages inferred 
for several of the very few systems 
with detected and measured gas content are relatively old -- 5-10 Myr!

\section{Closing in on the Future}

One point to emphasize, in particular, regarding circumstellar disk
evolution is that whatever happens to the early stage gas and dust disks, 
happens fast.  An increasingly important limit to our understanding
is thus the large uncertainty in stellar ages which leads 
to large uncertainty in disk evolution times, 
since $\Delta \tau_{evolution} < \delta \tau_{age}$.  In other words, 
the phenomenon occurs on time scales comparable to or less than
a stellar age resolution element.  Not so much later in the disk evolution
process, as primordial disks are dissipating, it is likely that
debris disks are arising. For these, now gas-poor disks, in contrast 
to the gas-rich primordial disk situation, 
$\Delta \tau_{evolution} > \delta \tau_{age}$.  In other words, 
the phenomenon occurs on somewhat longer time scales than
the uncertainties in stellar ages. 
Also needed for progress on disk evolution is semantic agreement 
on, and a common definition for, ``age," 
especially concerning a meaningful time zero point. 

Better understanding is needed of 
the transition from ``primordial" disks --
in which the mean grain size is increasing with time as sub-micron
material agglomerates to eventually form planets 
-- and ``debris" disks -- in which the mean grain size 
is decreasing with time as micron to cm-sized and larger material 
is destroyed in the planet-induced stirring of planetesimals 
and resulting collisional cascade, and followed by radiative blowout 
or inward drag.  Because the youngest examples of debris disks
overlap in age with the oldest known (accreting, even!) primordial disks,
we are going to have to be somewhat careful in parsing the data
in the 3-15 Myr age range where vestiges of both types of disk are likely present. 

As an example, Carpenter et al. 2006 may be seeing in their 5 Myr old sample
evidence for debris disks surrounding some of the earlier type (A-F) stars 
and unevolved primordial disks surrounding the later type (K-M) stars.  
It has also been argued by Metchev et al. 2005 in the case of 
the $\sim 12$ Myr old M-type star AU Mic, that while the inner regions
of the spatially resolved disk are collisionally evolved debris,
the outer regions are most likely pristine material that is still part
of the remnant primordial disk.  
The formation of debris disks, commonly accepted 
as evidence of formed planetesimals and even planets, 
can easily be confused with long-lived primordial disks.
In the absence or neglect of relevant information,
extreme caution and probably cleverness in the interpretation 
of observational data is needed. 
Gas studies are extremely promising in this regard.

Finally, we note that the relationship between disk evolution and
planet formation is becoming increasingly clear. We are currently in a stage 
of great luxury -- based on the substantial progress over the past
several years -- in being able refine the questions we can both
fathom and afford to ask of increasingly predictive theory and 
increasingly detailed observations.


\References
\item[]   
Alexander, R.D. \& Armitage, P.J. 2005, ApJL, 639, 83
\item[]   
Alexander, R.D. \& Armitage, P.J. 2007, MNRAS, 375, 500
\item[]   
Andrews, S.M. \& Williams, J.P. 2005, ApJ, 631, 1134 
\item[]   
Andrews, S.M. \& Williams, J.P. 2007, ApJ, 671, 1800
\item[]
Bary, J.S., Weintraub, D.A. \& Kastner, J.H. 2002, ApJL, 576, 73
\item[]
Bary, J.S., Weintraub, D.A. \& Kastner, J.H. 2003, ApJ, 586, 1136
\item[]   
Bertout, C., Siess, L. \& Cabrit, S. 2007, A\&AL, 473, 21
\item[]   
Bitner, M.A., Richter, M.J., Lacy, J.H., Greathouse, T.K., Jaffe, D.T., \& Blake, G.A.  2007, ApJL, 661, 69 
\item[]   
Brauer,  F., Dullemond, C.P. \& Henning, Th. 2008, A\&A 480, 859
\item[]   
Calvet, N. Briceno, C., Hernandez, J., Hoyer, S., Hartmann, L., Sicilia-Aguilar, A., Megeath, S.T., D'Alessio, P. 2005, AJ, 129, 935 
\item[]
Carpenter, J.M., Mamajek, E.E., Hillenbrand, L.A., \& Meyer, M.R. 2006, ApJL, 651, 49
\item[]
Carr, J. \& Najita, J.R. 2008, Science, 319, 1504
\item[]
Cieza, L. et al. 2007, ApJ, 667, 308 
\item[]
Dahm, S.E. \& Hillenbrand, L.A. 2007, AJ, 133, 2072
\item[]
Dahm, S.E. 2005, PhD Thesis, University of Hawaii
\item[]
Draine et al. 2006, ApJ, 636, 1114 
\item[]
Dutrey, A., Guilloteau, S., Duvert, G., Prato, L., Simon, M., Schuster, K., \& Menard, F. 1996, A\&A, 309, 493
\item[]
Edwards, S. et al. 1994, AJ, 108, 1056
\item[]
Furlan, E. et al. 2006, ApJS, 165, 568
\item[]
Garcia Lopez, R., Natta, A., Testi, L., \& Habart, E., 2006, A\&A, 459, 797
\item[]
Gorti, U. \& Hollenbach, D. 2008, ApJ, in press (arXiv0804.3381)
\item[]
Gullbring, E., Hartmann, L., Briceno, C., \& Calvet, N. 1998, ApJ, 492, 323
\item[]
Hartigan, P., Edwards, S., \& Ghandour, L., 1995, ApJ, 452, 736
\item[]
Hartmann, L., Calvet, N., Gullbring, E. \& D'Alessio, P. 1998, ApJ, 495, 385
\item[]
Hartmann, L. et al. 2005, ApJ, 629, 881 
\item[]
Herczeg, G.J., Najita, J., Hillenbrand, L.A., \& Pascucci, I. 2007, 670, 509
\item[]
Hernandez, J. et al. 2007, ApJ, 662, 1067 
\item[]   
Hillenbrand, L.A., 2005, review article to appear in "A Decade of Discovery: Planets Around Other Stars" STScI Symposium Series 19, ed. M. Livio (astro-ph/0511083)
\item[]
Hillenbrand, L.A. \& White, R.J. 2004, ApJ, 604, 741
\item[]
Hillenbrand, L.A., Bauermeister, A., \& White, R.J. 2009, in preparation
\item[]
Hillenbrand, L.A., Bauermeister, A., \& White, R.J. 2008, ASP Conf. Ser. 384, 200
\item[]   
Hirota, T., Bushimata, T., Choi, Y.K. et al., 2007, PASJ, 60, 37  
\item[]
Hollenbach, D., Johnstone, D., Lizano, S., \& Shu, F. 1994, ApJ, 428, 654
\item[]
Jayawardhana, R., Coffey, Jaime, Scholz, A., Brandeker, A., \& van Kerkwijk, M.H., 2006, ApJ, 648, 1206
\item[]   
Jeffries, R.D., 2007, MRNAS, 376, 1109
\item[]
Lada, C.J. et al. 2006, AJ, 131, 1574
\item[]
Lahuis, F., van Dishoeck, E.F., Blake, G.A., Evans, N.J., II., Kessler-Silacci, J.E., \& Pontoppidan, K.M. 2007, ApJ, 665, 492
\item[]
Low, F.J., Smith, P.S., Werner, M., Chen, C., Krause, V., Jura, M., Hines, D.C. 2005, ApJ, 631, 1170
\item[]
Mamajek, E.E. \& Hillenbrand, L.A. 2008, ApJ, submitted
\item[]
Mamajek, E.E., Meyer, M.R., Hinz, P.M., Hoffmann, W.F., Cohen, M., \& Hora, J.L. 2004, ApJ, 612, 496
\item[]
Megeath, S.T., Hartmann, L., Luhman, K.L., \& Fazio, G.G. 2005, ApJL, 634, 113  
\item[]   
Menten, K.M., Reid, M.J., Forbrich, J., \& Brunthaler, A. 2008, A\&A, 474, 515
\item[]
Metchev, S.A., Eisner, J.A., Hillenbrand, L.A., Wolf, S. 2005, ApJ, 622, 451
\item[]
Mohanty, S., Jayawardhana, R., \& Basri, G., 2005, ApJ, 626, 498
\item[]
Monin, J.-L., Clarke, C.J., Prato, L. \& McCabe, C. 2007, in Protostars and Planets V, ed. B. Reipurth, D. Jewitt, \& K. Keil (Tucson: Univ. Arizona Press), 395
\item[]
Muzerolle, J., Hillenbrand, L., Calvet, N., Briceno, C. \& Hartmann, L. 2003, ApJ, 592, 266 
\item[]
Najita, J.R. \& Strom, S.E. \& Muzerolle, J. 2007, MNRAS, 378, 369
\item[]
Nordh, L. et al. 1996, A\&AL, 315, 185
\item[]
Osterloh, M. \& Beckwith, S.V.W. 1995, ApJ, 439, 288
\item[]
Pascucci, I. et al. 2007, ApJ, 663, 383
\item[]
Pudritz, R.E., Ouyed, R., Fendt, Ch. \& Brandenburg, A. 2007, in Protostars and Planets V, ed. B. Reipurth, D. Jewitt, \& K. Keil (Tucson: Univ. Arizona Press), 277
\item[]
Ramsay Howat, S.K. \& Greaves, J.S., 2007, MNRAS, 379, 1658
\item[]
Salyk, C., Blake, G.A., Boogert, A.C.C. \& Brown, J.M. 2007, ApJL, 655, 105
\item[]   
Salyk, C, Pontoppidan, K.M., Blake, G.A., Lahuis, F.,
van Dishoeck, E.F., \& N.J. Evans II, 2008, ApJL, 676, 49
\item[]
Sandstrom, K.M., Peek, J.E.G., Bower, G.C. et al. 2007, ApJ, 667, 1161
\item[]
Sicilia-Aguilar, A., Hartmann, L., Furesz, G., Henning, T., Dullemond, C. \& Brandner, W.  2006, AJ, 132, 2135 
\item[]
Sicilia-Aguilar, A. et al. 2006, ApJ, 638, 897
\item[]
Silverstone, M.D. et al. 2006, ApJ, 639, 1138
\item[]
Simon, M. \& Prato, L. 1995, ApJ, 450, 824
\item[]
Slesnick, C.L., Hillenbrand, L.A. \& Carpenter, J.M., 2008, ApJ, submitted. 
\item[]
Skrutskie, M.F., Dutkevitch, D., Strom, S.E., Edwards, S., Strom, K.M. \& Shure, M.A.  1990, AJ, 99, 1187
\item[]
Takeuchi, T., Clarke, C.J., \& Lin, D.N.C., 2005, ApJ, 627, 286
\item[]
Tout, C.A., Livio, M. \& Bonnell, I.A. 1999, MNRAS, 310, 360
\item[]
Watson, D.M. et al. 2007, arXiv/0704.1518 
\item[]
Weintraub, D.A., Kastner, J.H. \& Bary, J.S. 2000, ApJ, 541, 767
\item[]
White, R.J. \& Basri, G. 2003, ApJ, 582, 1109
\item[]
Wolk, S.J. \& Walter, F.M. 1996, AJ, 111, 2066

\endrefs

\end{document}